\documentclass[sn-mathphys,Numbered]{sn-jnl}

\usepackage{graphicx}%
\usepackage{multirow}%
\usepackage{amsmath,amssymb,amsfonts}%
\usepackage{amsthm}%
\usepackage{mathrsfs}%
\usepackage{physics}
\usepackage{lmodern}

\usepackage[title]{appendix}%
\usepackage{xcolor}%
\usepackage{textcomp}%
\usepackage{manyfoot}%
\usepackage{booktabs}%
\usepackage{algorithm}%
\usepackage{algorithmicx}%
\usepackage{algpseudocode}%
\usepackage{listings}%

\providecommand{\vc}{\vb*}

\renewcommand{\Re}{\real}
\renewcommand{\Im}{\imaginary}

\begin{document}

\title[]{Longitudinal chiral forces in photonic integrated waveguides to separate particles with realistically small chirality}


\author[1]{\fnm{Josep} \sur{Martínez-Romeu}}
\author[1]{\fnm{Iago} \sur{Diez}}

\author[2]{\fnm{Sebastian} \sur{Golat}}

\author[2]{\fnm{Francisco J.} \sur{Rodríguez-Fortuño}}

\author[1]{\fnm{Alejandro} \sur{Martínez}} \email{amartinez@ntc.upv.es}

\affil[1]{\orgdiv{Nanophotonics Technology Center}, \orgname{Universitat Politècnica de València}, \orgaddress{\street{Camino de Vera, s/n Building 8F}, \postcode{46022}, \city{Valencia}, \country{Spain}}}

\affil[2]{\orgdiv{Department of Physics}, \orgname{King's College London}, \orgaddress{\street{Strand}, \postcode{WC2R 2LS}, \city{London},  \country{United Kingdom}}}





\abstract{\textbf{} Chiral optical forces exhibit opposite signs for the two enantiomeric versions of a chiral molecule or particle. If large enough, these forces might be able to separate enantiomers all optically, which would find numerous applications in different fields, from pharmacology to chemistry. Longitudinal chiral forces are especially promising for tackling the challenging scenario of separating particles of realistically small chiralities. In this work, we study the longitudinal chiral forces arising in dielectric integrated waveguides when the quasi-TE and quasi-TM modes are combined as well as their application to separate absorbing and non-absorbing chiral particles. 
We show that chiral gradient forces dominate in the scenario of beating of non-denegerate TE and TM modes when considering non-absorbing particles. For absorbing particles, the superposition of degenerate TE and TM modes can lead to chiral forces that are kept along the whole waveguide length. We accompany the calculations of the forces with particle tracking simulations for specific radii and chirality parameters. We show that longitudinal forces can separate non-absorbing chiral nanoparticles in water even for relatively low values of the particle chirality and absorbing particles with arbitrarily low values of chirality can be effectively separated after enough interaction time.}





\maketitle  
\clearpage
\section{Introduction}


Chirality is a property of asymmetry of objects which holds great importance in different branches of science and technology. This property describes objects that cannot be superimposed with their mirrored selves and is present from subatomic particles to macroscopic structures. This includes the remarkable case of molecular chirality \cite{Barron_2004}, by which molecules can display two opposite handedness, the so-called right- and left-handed enantiomers, which may show completely different physical and chemical properties. As an example in medicine, one molecule can have medicinal properties while its opposite enantiomer can be extremely toxic \cite{Jacques1981}. Therefore, it is of utmost importance to be able to separate the two types of enantiomers of certain chemical substances with great accuracy, in great volumes, and in a short time. 

Usual methods for separating enantiomers from mixtures rely on chemical processes that must be changed for each specific molecule. Alternatively, one could take advantage of the electromagnetic properties of chiral molecules, which have been thoroughly studied \cite{Kuhn,Mason63}, and use the chiral optical forces exerted by light \cite{Genet2022,BLIOKH20151,Bliokh2014,Zhao2017,Schnoering2021,Magallanes2018,Kivshar2014} to perform enantiomer separation. Remarkably, this interaction does not depend on the specific molecule, which presents a great advantage over chemically-based separation. 

Due to the prospects for application in different industries, optical separation of enantiomers has recently received considerable attention, including many theoretical and simulation studies \cite{Canaguier-Durand2013, Hayat2015,Canaguier-Durand2015,Zhang2017,Cao2019,Zheng2020,Zhang2021,Wang2014} as well as some experimental implementations \cite{Tkachenko2014,Magallanes2018,Tkachenko20141,Shi2020}. Most previous works have considered free-space light beams incident upon chiral structures so that separation forces are exerted locally. A different approach proposes the use of guided light along dielectric fibers \cite{Golat2023} or integrated waveguides \cite{Martinez-Romeu2024} to exert transverse chiral forces over long (in terms of wavelength) propagation lengths that eventually could lead to enantiomeric sorting. However, such forces usually require large values of the chirality parameter of the nanoparticles to lead to partial separation, meaning that other strategies are needed to separate nanoparticles and molecules exhibiting lower chiral response, as usually happens in practice. 

In this work, we circumvent this problem by using longitudinal chiral forces arising in dielectric integrated waveguides upon the superposition of the two fundamental guided modes: the quasi-TE and the quasi-TM modes. We consider the cases of both absorbing and non-absorbing chiral nanoparticles. 
For non-absorbing nanoparticles, we show that in a waveguiding system where the electromagnetic energy density does not vary but the field helicity does vary, low chirality particles could be separated. A similar approach was followed for free-space optical beams using diffraction gratings and reflection in a gap between a prism and substrate \cite{Cameron2014_Discriminatory,Cameron2014,Yao2024}. In our guided approach, we show that a photonic integrated waveguide can be designed to produce longitudinal optical chiral forces stronger than the achiral forces for a wide range of both the particle radius and the chirality parameter. Our numerical results suggest that such photonic waveguide could lead to enantiomeric sorting along the propagation direction of light.

For absorbing particles, we change the approach we use because the dominant forces will change. In this case, we leave behind the idea of having stronger chiral forces than achiral forces and focus on maintaining a chiral force over a long waveguide, which can be achieved by using guided chiral light making use of degenerate and $90^\circ$-shifted quasi-TE and quasi-TM modes. We show that separation is feasible along the longitudinal direction even in the case of small chirality of the nanoparticle.

\section{Review of optical chiral forces} \label{Forces}

Optical forces exerted by an optical field on a chiral dipolar particle have been thoroughly studied in the literature \cite{Genet2022,Golat2023}. We choose the particle-centric form of the force expression that is described in detail in previous work \cite{Golat2023}:
%
\begin{equation}\label{eq:all_forces}
    \begin{split}
    \vc{F}=&\underbrace{\grad(\Re\alpha_\text{e}W_\text{e}\!+\!\Re\alpha_\text{m}W_\text{m}\!+\!\Re\alpha_\text{c}\omega\mathfrak{G}\! )}_\text{gradient force}
    \\&\underbrace{+2\omega(\Im\alpha_\text{e}\vc{p}_\text{e}\!+\!\Im\alpha_\text{m}\vc{p}_\text{m}\!+\!\Im\alpha_\text{c}\Re\vc{p}_\text{c}\! )}_\text{radiation pressure force}
    \\&\underbrace{-(\sigma_\text{rec}\Re\vc{\varPi}\!+\!\sigma_{\text{im}}\Im\vc\varPi)/c
    -\omega(\gamma^\text{e}_\text{rec} \vc{S}_\text{e}\!+\!\gamma^\text{m}_\text{rec} \vc{S}_\text{m})}_\text{dipole recoil force}.\!\!\!
\end{split}
\end{equation}

\noindent where $\omega$ is the angular frequency, $k$ is the wavenumber, $W_{\rm e} = \frac{1}{4}\varepsilon |\vc{E}|^2$ and
$W_{\rm m} = \frac{1}{4}\mu |\vc{H}|^2$ are the electric and magnetic energy densities, respectively, measured in $ \left[\mathrm{J}/ \mathrm{m}^3\right]$ units.
The helicity density is $\mathfrak{G} = \frac{1}{2\omega c} \imaginary \left( \vc{E} \cdot \vc{H}^* \right)$ $ \left[\mathrm{J}\cdot \mathrm{s}/\mathrm{m}^3\right]$, whose sign indicates the handedness of the optical field. The following field properties $\vc{S}_{\rm e} = \frac{1}{4\omega} \imaginary \left( \varepsilon \ \vc{E}^*\, \times \vc{E} \right) $ $\left[\mathrm{J}\cdot \mathrm{s}/\mathrm{m}^3\right]$ and $
\vc{S}_{\rm m} = \frac{1}{4\omega} \imaginary \left(\mu \, \vc{H}^*\times \vc{H} \right) $ $\left[\mathrm{J}\cdot \mathrm{s}/\mathrm{m}^3\right]$ yield respectively the electric magnetic spin densities of the field. The complex Poynting vector is represented by $\vc{\varPi} = \frac{1}{2}\vc{E} \times \vc{H}^* $ $\left[\mathrm{W}/\mathrm{m}^2\right]$. The electric, magnetic and chiral momentum of the light field are respectively   
$\vc{p}_\text{e}=\frac{1}{2c^2}\Re\vc{\varPi}-\frac{1}{2}\curl{\vc{S}_\text{e}}$, $\vc{p}_\text{m}=\frac{1}{2c^2}\Re\vc{\varPi}-\frac{1}{2}\curl{\vc{S}_\text{m}}$ and $\Re\vc{p}_\text{c}=k(\vc{S}_\text{e}+\vc{S}_\text{m})-\frac{1}{2\omega c}\curl{\Re\vc{\varPi}}$ \cite{Golat2023,Berry2009,Vernon2023}. 

The properties of the particle are characterized by the electric polarizability $\alpha_{\rm e}$, the magnetic polarizability $\alpha_{\rm m}$, and the chiral polarizability $\alpha_{c}$. The latter informs about how electromagnetically chiral the particle is. The other constants depend on the product of polarizabilities, $\sigma_\text{rec}=\frac{k^4}{6 \pi}[\Re(\alpha_{\text{e}}^* \alpha_{\text{m}})+\left|\alpha_{\text{c}}\right|^2]$, $\sigma_\text{im}=\frac{k^4}{6 \pi}\Im(\alpha_{\text{e}}^* \alpha_{\text{m}})$, $\gamma^\text{e}_\text{rec}=\frac{k^4}{3\pi }\Re(\alpha_\text{e}^\ast\alpha_\text{c})$, $\gamma^\text{m}_\text{rec}=\frac{k^4}{3\pi }\Re(\alpha_\text{m}^\ast\alpha_\text{c})$. Even though the full expression of the forces is used to calculate the total force, it is important to know which terms dominate to gain insight into the physics of the system. In particular, we will consider first the separation of non-absorbing chiral particles, followed by the separation of absorbing chiral particles. In the former scenario, the dominant forces are those relying on the real part of the polarizabilities (gradient terms), while in the latter the dominant terms will be those proportional to the imaginary part of the polarizabilities (radiation pressure terms). Another important aspect is that for small particles the dominant forces will be those that depend on the polarizabilities up to the first order. 

To study the forces acting on the chiral particle, we need to know the electric and magnetic field profile of the quasi-TE and quasi-TM guided modes of the waveguide (eigenmodes). First, the two-dimensional profile of the fields ($\vc{E}(x,y)$, $\vc{H}(x,y)$) is obtained by solving Maxwell's equations throughout the two-dimensional cross-section of the waveguide system using the finite element method implemented by the FemSIM solver in the commercial software RSoft (Synopsis). Then, these profiles are propagated along the waveguide using the corresponding effective indices $n_{\rm TE}$ for the TE mode and $n_{\rm TM}$ for the TM mode. For instance, for the electric field of the quasi-TE: $\vc{E}(x,y,z)=\vc{E}(x,y)e^{ikn_{\rm TE}z}$. This propagation results in a three-dimensional field along the waveguide. The resulting fields are then inserted into Eq.~\ref{eq:all_forces} to compute the force exerted by the mode on a dipolar particle \cite{Golat2023}. Chiral Mie theory \cite{Bohren1975} was used to calculate the polarizabilities of the particle from the properties of the particle ($r$, $\varepsilon_{\rm p}$, $\mu_{\rm p}$, $\kappa$) and the surrounding medium ($\varepsilon_{\rm m}$, $\mu_{\rm m}$). The chirality parameter $\kappa$ characterizes the difference of the refractive index for left circularly polarized light and right circularly polarized light traveling through an optically active medium. For a medium constituted of $(+)$ or $(-)$ enantiomers the refractive index is: $n_{\pm}=n\pm \kappa$. This parameter is then particularly important to the study of chirality. $\Re(\kappa)$ is related to the optical rotatory dispersion and $\Im(\kappa)$ is related to circular dichroism \cite{Bohren1975}.

\begin{figure}[t]
    \centering
    \includegraphics[width=0.95\textwidth]{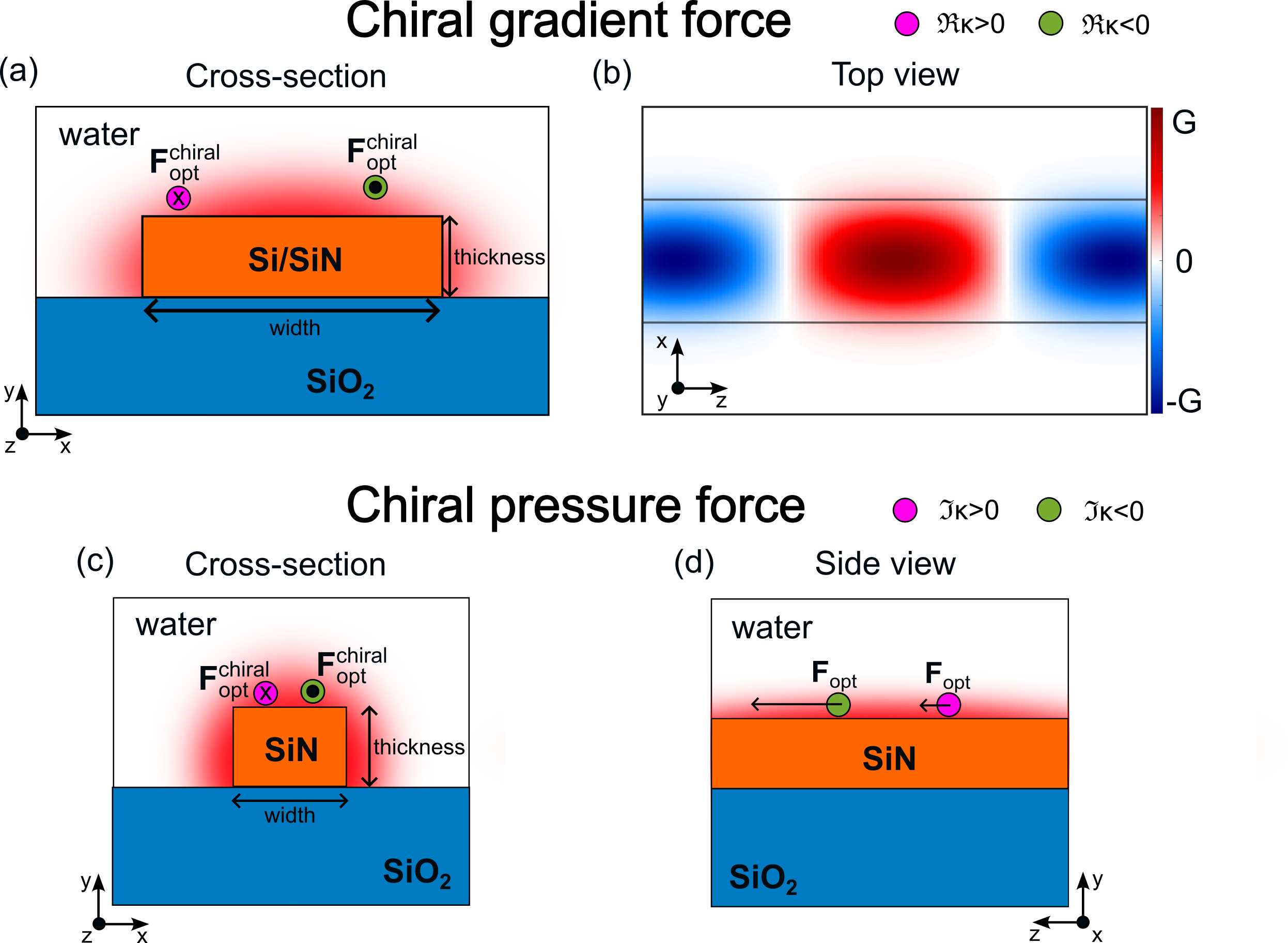}
    \caption{Conceptual representation longitudinal chiral forces in integrated waveguides. (a) Cross-section schematic of a wide dielectric waveguide representing the opposite action of the total optical force exerted by the mode in the waveguide (represented in fade red) onto particles with opposite chirality. The x or $\cdot$ mark on the particle represents the direction of the force along the $z$ axis (the force points into the paper or out of the paper, respectively).  (b) Representation of helicity density over a waveguide where a TE mode and a TM mode phase-shifted 90 degrees have been injected. The helicity flips sign every half a beat length, where the beat length is defined as $L_{\rm beat}=\lambda / (n_{\rm TE}-n_{\rm TM})$. (c) Schematic of a waveguide supporting degenerate quasi-TE and quasi-TM modes. The sorting mechanism consists of a different velocity of chiral particles. The x or $\cdot$ mark on the particle represents the opposite direction of the force in the $z$ axis for the opposite enantiomers. (d) Side view of the waveguide showing that the chiral particles are pushed towards the same direction but with a different total force depending on their handedness.
    }
    \label{fig:Description_system}
\end{figure}

\section{Longitudinal chiral gradient force on non-absorbing nanoparticles} \label{Gradient Forces}
We first consider non-absorbing particles, i.e. real $\varepsilon_{\rm p}$ and real $\kappa$, the dominant force terms are the gradient forces, which are proportional to the real part of the polarizabilities. Notably, if the particles are large enough the Poynting part of the recoil force must be taken into account. The resulting force can be approximated as:
\begin{equation}\label{eq:dominant_forces}
    \vc{F} \approx \omega \; \real\alpha_{\rm c} \nabla \mathfrak{G} \; \; + \; \; {\real\alpha_{\rm e}\nabla W_{\rm e}} \; \; + \; \; {\real\alpha_{\rm m}\nabla W_{\rm m}-\sigma_\text{rec}\Re\vc{\varPi}/c}
\end{equation}  

From the previous equation, we can conclude that a system where the electric and magnetic energy density gradients (achiral forces) are negligible in comparison with the helicity density gradient (chiral force) should be in principle able to produce enantioseparation. This condition can be met along the longitudinal direction for a lossless dielectric integrated waveguide with rectangular cross-section. In these waveguides, the light intensity of the mode is maintained over the longitudinal direction. This means that the electric and magnetic energy densities do not change, i.e. the gradient of both energy densities (and therefore the associated achiral forces) are negligible in the longitudinal direction regardless of the electric and magnetic polarizability of the particle. To produce enantioseparation, the waveguiding system should present a chiral force 
(helicity density gradient), given by the first term of Eq. \ref{eq:dominant_forces}. This helicity density gradient can be produced within the waveguide by achieving a guided wave whose helicity density varies longitudinally, thus producing a helicity gradient and, therefore, a chiral longitudinal force. This setup can be achieved by simultaneously injecting the fundamental quasi-TE and quasi-TM modes in a waveguide where $n_{\rm TE}\neq n_{\rm TM}$.A schematic of such a waveguide is depicted in Fig. \ref{fig:Description_system}a whilst the helicity change in the longitudinal direction is shown in Fig. \ref{fig:Description_system}b.
The longitudinal component of the helicity density gradient is given by the following equation (analytical derivation in section \ref{app_derivation} of the appendix):
\begin{equation}\label{eq:helicity_grad_force}
    F_z^{\nabla\mathfrak{G}}= \Re\alpha_{\rm c} \omega\frac{d\mathfrak{G}}{dz} = \frac{1}{4 c}\Re\alpha_{\rm c}\left|\psi\right| k\Delta n \sin(k\Delta n z + \arg\psi)
\end{equation}

\noindent where $\psi=\vc{E}_{\rm TE}\cdot \vc{H}_{\rm TM}^*-\vc{E}_{\rm TM}^*\cdot \vc{H}_{\rm TE}$. Equation~\ref{eq:helicity_grad_force} shows that the chiral force oscillates sinusoidally with a periodicity of $L_{\rm beat}=\lambda/\Delta n$, which is referred to as the beat length, and flips sign every half of the beat length $L_{\rm beat}/2=\lambda/(2\Delta n)$, where $\Delta n=n_{\rm TE}-n_{\rm TM}$ is the difference between the TE and TM mode indices. Therefore, by injecting left-handed elliptically polarized (LEP) light, which can be achieved by a $90\circ$-phase-shifted combination of the quasi-TE and quasi-TM guided modes, at the waveguide input, the polarization of the guided wave will change to right-handed elliptically polarized (REP) guided light after half of the beat length. The dominant chiral longitudinal force is proportional to the longitudinal gradient of the helicity density ($F_z=\omega \real\alpha_{\rm c}\nabla_z \mathfrak{G}$) and, therefore, it is the force enabling the enantioseparation. This force increases with $\Delta n$ and decreases with the wavelength and will always be present whenever mixing two non-degenerate guided modes. The separated distance of the enantiomers is inversely proportional to $k\Delta n$. Therefore changing the $k\Delta n$ to augment the forces must be balanced with keeping enough distance between the enantiomers to produce significant separation. For small particles, the helicity density gradient force is the only dominant force term in the longitudinal direction, thus $F_{z,\rm achiral}<F_{z,\rm chiral}$, which favors enantiomeric separation even for particles exhibiting low chirality ($\real \alpha_{\rm c}\ll\real\alpha_{\rm e},\real\alpha_{\rm m}$).




\section{Longitudinal chiral pressure force on absorbing nanoparticles} \label{Pressure Forces}

For absorbing particles, i.e. complex $\varepsilon_{\rm p}$ and purely imaginary $\kappa$, 
we will consider a waveguiding system where a quasi-circularly polarized mode is injected by combining the quasi-TE mode and 90$^\circ$-delayed quasi-TM mode that are degenerate ($n_{\rm TE}=n_{\rm TM}$). In this case, there will be no chiral or achiral gradient forces in the longitudinal direction since the wave helicity is conserved along the propagation direction.  
Moreover, if the particles are small enough, the dominant forces will be those that depend linearly on the polarizabilities, which are exclusively the radiation pressure forces from Eq.~\ref{eq:all_forces}. Taking this into account, the resulting dominant longitudinal optical forces for small particles are:
\begin{equation}\label{eq:dominant_forces2}
{F_z} \approx 2\omega(\Im\alpha_\text{e}{p}_{\text{e},z}\!+\!\Im\alpha_\text{m}{p}_{\text{m},z}\!+\!\Im\alpha_\text{c}\Re{p}_{\text{c},z})
\end{equation}  

Notably, the electric and magnetic pressure forces do not change along the length of the waveguide as long as it is lossless. As a result, both enantiomers will be equally pushed in the longitudinal direction. The difference in the force exerted upon the two enantiomers will be due to the chiral pressure, which depends on the longitudinal spin of the light. Since the quasi-TE and quasi-TM modes do not display longitudinal spin \cite{Espinosa-Soria2016}, we need a suitable combination of them to get longitudinal spin: they must have equal amplitude and be $90\circ$ phase shifted at the operating wavelength. Moreover, they need to exhibit degeneracy ($n_{\rm TE}=n_{\rm TM}$), which can be achieved by proper design of the waveguide cross-section, so that the longitudinal spin is maintained along the length of the waveguide \cite{Vázquez‐Lozano2020}. 

For absorbing particles, $F_{\rm achiral}>F_{\rm chiral}$ holds for a wide range of small chiralities, unlike in the non-absorbing chiral-beating setup. This condition implies that there can only be enantioseparation if we do not have a trapping achiral force. In our case, the achiral force pushes the particles along the waveguide, meaning the difference of forces can be used to separate the racemic mixture. For small chirality, the chiral force will be quite small in magnitude but, as we explored in a previous work \cite{Martinez-Romeu2024}, even if the separating chiral force is small, it will eventually overcome the Brownian motion and separate the racemic mixture if applied for long enough time. As a result, an important aspect of this approach is maintaining the chiral force over a long distance, which can be achieved via the degeneracy of the main guided modes. This sorting mechanism is schematized in Fig.~\ref{fig:Description_system}c showing the opposite direction of the chiral part of the force for opposite enantiomers. Figure~\ref{fig:Description_system}d shows that despite the chiral part of the force is smaller than the achiral part, the particles will acquire different velocities that will eventually lead to their separation.

\section{Results} 


\subsection{Chiral gradient forces}
We consider a strip waveguide made of a silicon core (0.480 $\mu $m width $\times$ 0.220 $\mu $m thickness) with refractive index $n\approx3.45$ on a SiO$_2$ substrate $(n=1.4468)$, surrounded by water, operating at a wavelength of 1310 nm. The high index contrast between Si and SiO$_2$/water was chosen to increase $\Delta n$, and thus, the enantioseparating force as shown by Eq.\ref{eq:helicity_grad_force}. Widening the waveguide also increases $\Delta n$; however as the guided power spreads over a larger area, the local energy and helicity densities become smaller, and, consequently, the forces also diminish. For comparison, we also consider a silicon nitride ($n\approx2$) waveguide (1.170 $\mu$m width $\times$ 0.268 $\mu$m thickness) operating at shorter wavelengths (780 nm) at which silicon becomes absorbing. In both cases (silicon and silicon nitride), we look for the same effects: gradient chiral forces being stronger than the achiral forces for different radii and chirality parameters. To this end, we consider non-absorbing chiral particles with relative permittivity $\varepsilon_{\rm p} = 2$ and relative permeability $\mu=1$, suspended in water ($n=1.33$). The studied waveguide cross-section and the beating of helicity density along the longitudinal direction above the waveguide are shown in Fig.~\ref{fig:Description_system}(a) and (b), respectively.



\begin{figure}[t]
    \centering
    \includegraphics[width=1\textwidth]{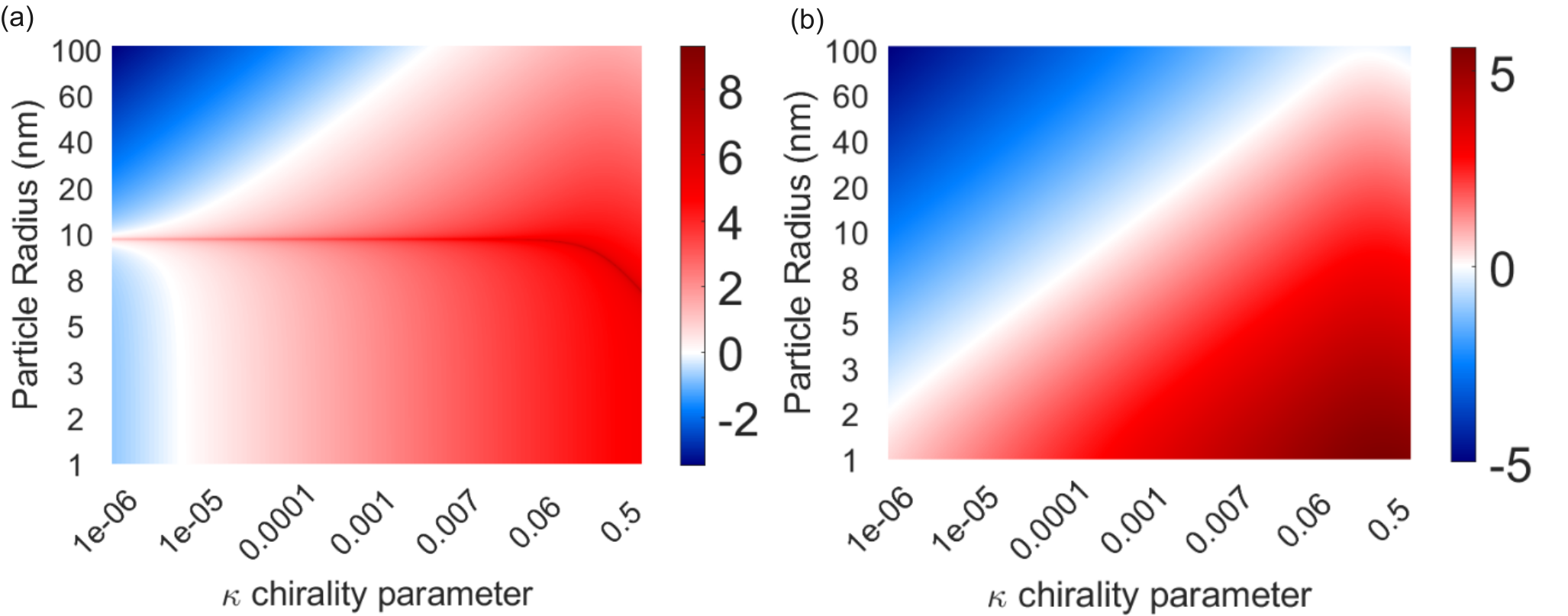}
    \caption{Gradient chiral vs. achiral forces in the two waveguide systems: (a) silicon waveguide (0.480 $\mu$m wide $\times$ 0.220 $\mu$m thick) operating at 1310 nm wavelength; (b) silicon nitride waveguide (1.170 $\mu$m wide $\times$ 0.268 $\mu$m thick), operating at 780 nm wavelength. The colormaps represent $\log_{10}(|F^{\rm chiral}_{z}/F^{\rm achiral}_{z}|)$ for different combinations of particle radius and chirality parameter. In the red zones $|F^{\rm achiral}_{z}|<|F^{\rm chiral}_{z}|$, in the blue zones $|F^{\rm achiral}_{z}|>|F^{\rm chiral}_{z}|$, and in the white zones the chiral and achiral forces show similar strength.  The coordinates ($x,y,z$) belong to the same axis as shown in Fig.\ref{fig:Description_system}. The ratio was evaluated at a centered position in $x$, 0.1 $\mu$m over the top of the waveguide in $y$, and at a $z$ where the point of helicity sign change: $0.53$ $\mu$m for (a) and 2 $\mu$m for (b).}
    \label{fig:Radii_kappa_sweep}
\end{figure}

We first studied which combinations of particle size and chirality parameter favor enantioseparation in this waveguide system. To this end, we plotted the ratio of the longitudinal total chiral force and the longitudinal total achiral force ($|F^{\rm chiral}_{z}/F^{\rm achiral}_{z}|$) in Fig.~\ref{fig:Radii_kappa_sweep} as a colormap, against the different chirality parameter and radii. The red (blue) zones represent where the longitudinal chiral forces are stronger (weaker) than the longitudinal achiral forces throughout the parameter space. The white zone shows where the chiral and achiral forces have a similar magnitude. Figure~\ref{fig:Radii_kappa_sweep}a corresponds to the silicon waveguide system and Fig.~\ref{fig:Radii_kappa_sweep}b to the silicon nitride waveguide system, both evaluated at the point of the helicity sign change.

The most important forces in the system are $F^{\nabla \mathfrak{G}}_{z}$, $F^{\nabla E}_{z}$ and $F^{\Re\Pi}_{z}$, and they can be combined in the parameter space to produce different regions of dominance of the achiral and chiral forces.  In Fig. \ref{fig:Radii_kappa_sweep}a, we observe three distinct regions favorable for separation of enantiomers (red zones): I) $\kappa > 10^{-5}$ and $r<10$ nm, II) $r>10$ nm and $10^{-6}<\kappa<10^{-3}$, III) $r\approx 10$ nm. In Fig. \ref{fig:Radii_kappa_sweep}b, we find only two distinct regions. To further explain how these forces interact in the parameter space to produce the resulting total force, we analyze the individual contribution of the force terms $F^{\nabla \mathfrak{G}}_{z}$, $F^{\nabla E}_{z}$ and $F^{\Re\Pi}_{z}$ for the silicon waveguide in section \ref{app_force_analysis} of the appendix. 

\begin{figure}[t]
    \centering
    \includegraphics[width=1\textwidth]{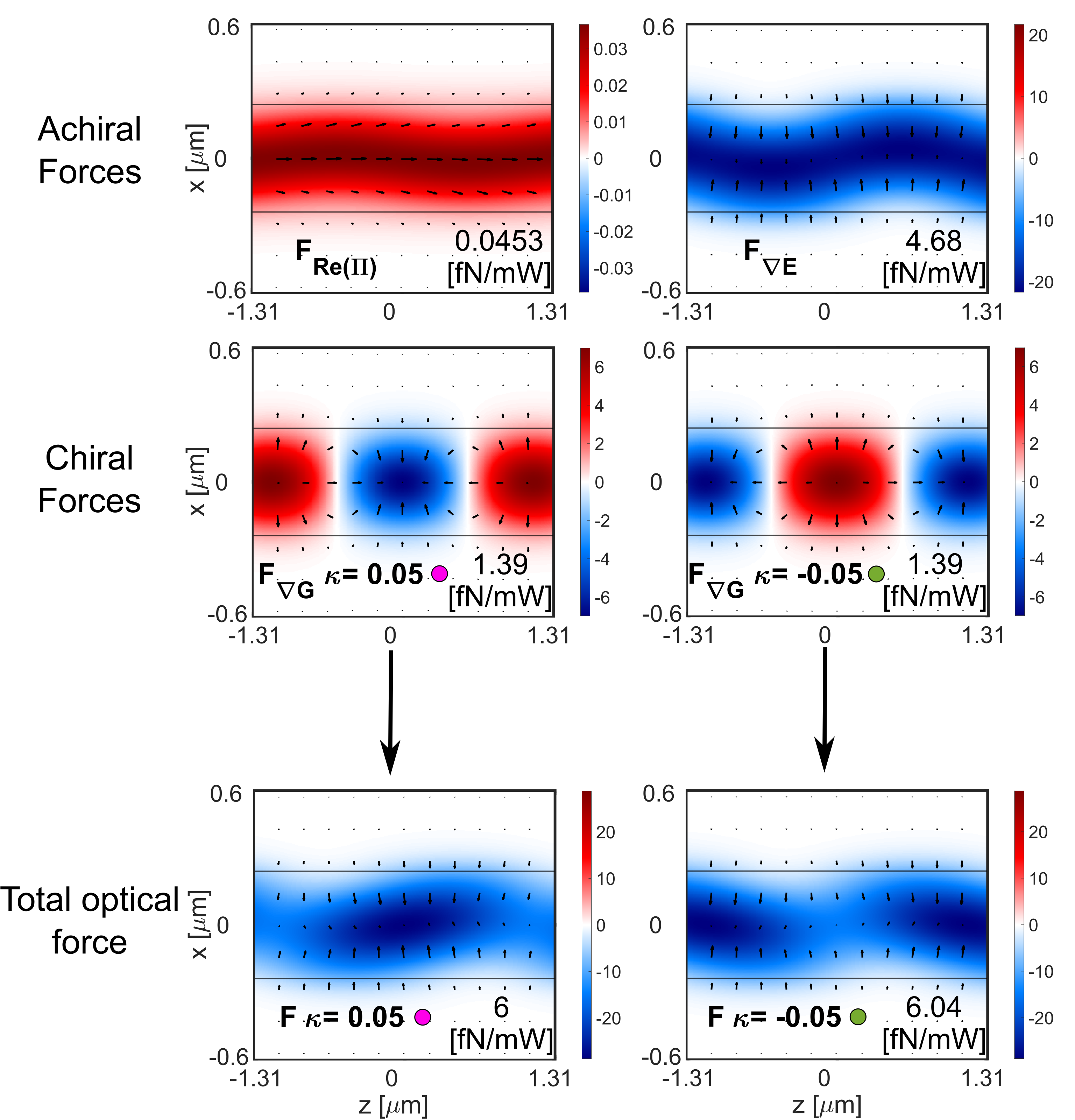}
    \caption{Optical forces exerted on each enantiomer ($r=100$ nm, $\kappa=\pm0.05$) by the optical mode propagating in the silicon waveguide (0.480 $\mu$m wide $\times$ 0.220 $\mu$m thick) operating at 1310 nm wavelength. We represent the most dominant chiral force and achiral forces as well as the total force.}
    \label{fig:All-Forces}
\end{figure}

We have chosen a combination of particle radius ($r=100$ nm) and chirality parameter ($\kappa=\pm 0.05$) that favors enantioseparation (from region II in Fig.~\ref{fig:Radii_kappa_sweep}a) to test the sorting capabilities of our designed silicon waveguide (width 0.480 $\mu$m $\times$ thickness 0.220 $\mu$m). For such a particle, the most dominant achiral ($F^{\nabla E}$ and $F^{\Re\Pi}$) and chiral forces ($F^{\nabla \mathfrak{G}}$) are shown in Fig.\ref{fig:All-Forces} throughout the $x-z$ plane situated 100 nm above the waveguide (same axis as in Fig. \ref{fig:Description_system}). The arrowmaps represent the force fields exerted on the particle by the guided field along the $x$ and $z$ directions. The colormaps represent the $y$-component of the force, which moves the particle towards (negative, in blue) or away (positive, in red) from the waveguide. The chiral gradient force has a maximum of 1.39 fN/mW in the x-z plane. The electric gradient force has a maximum of 4.68 fN/mW in the x-z plane. The total optical force pushes the particles towards the waveguide in height (represented in the colormap). The position where the particles are most atracted towards the waveguide in height changes with the chirality of the particle. This phenomenon is complementary to the chiral separation in the longitudinal direction. In the points along z where the opposed chiral particles are trapped, they have the highest attracting force towards the waveguide too.

The dominant force in both transversal directions $x$ and $y$ is the achiral electric energy density gradient $F^{\nabla E}$ that moves both enantiomers toward the center of the waveguide (in $x$) and toward the top of the waveguide (in $y$). At the center of the waveguide, $F^{\nabla E}$ exhibits a much smaller magnitude, and the chiral force $F^{\nabla \mathfrak{G}}$ dominates over both achiral forces ($F^{\nabla E}$ and $F^{\Re\Pi}$) along the longitudinal direction, thus enabling the sorting. This force analysis agrees with the selected particle radius and chirality parameter combination from Fig. \ref{fig:Radii_kappa_sweep}.

We have tested the sorting capability of this system by performing particle tracking simulations using the force field from Fig.~\ref{fig:All-Forces} for a guided power of 100 mW and during 1 second. The tracking algorithm is explained in detail in \cite{Martinez-Romeu2024}. For this type of simulation, we have assumed that the microchannel (where the particles are suspended in water) is placed perpendicular to the waveguide, i.e. along the $x-$direction with $12$ $\mu$m length, 1 $\mu$m width along $z-$axis, and 
 $0.5$ $\mu$m height along $y-$axis. Notice that the width along $z$ should be at least equal to half of the beat length to achieve enough separation distance. To obtain a statistical measurement of the success of the sorting process, we have conducted the individual tracking of each enantiomer 500 times. Each particle's starting position is randomized each time throughout an area of 400 nm $\times$ 400 nm in the $xz-$plane at 140 nm above the waveguide. 
 
\begin{figure}[t]
    \centering
    \includegraphics[width=1\textwidth]{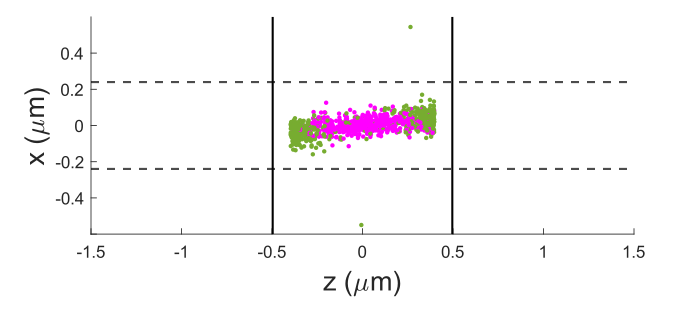}
    \caption{Result of the particle tracking simulation for 1 second and an optical power of 100 mW in the silicon waveguide of Fig \ref{fig:Description_system}. The final position of each enantiomer of the 500 chiral particles is represented by the magenta $(+)$ and green $(-)$ circles. The waveguide edge is represented by the dashed lines and the microchannel edge is represented by the solid lines. The $(+)$-particles are attracted towards the center of the microchannel whereas the $(-)$-particles are repelled, thus producing the separation.}
    \label{fig:Particle_Tracking}
\end{figure}

 The final positions of the particles are represented in Fig. \ref{fig:Particle_Tracking}, where we can see that 1 second is enough duration to sort both enantiomers. The ($+$)-enantiomer (in magenta) gets trapped in the center of the microchannel where the helicity density is positive (see Fig.\ref{fig:Description_system}b), whereas the ($-$)-particle is repelled from that zone and attracted towards the area where the helicity density is negative. 
The results show that 95\% of the $(+)$-particles end up within $z \in [-0.330, 0.330]$ $\mu$m, and 59.8\% ($-$)-particles end up outside. The latter percentage would be larger if the channel was wider. The purity of the mixture within a region is calculated with the quantity named enantiomeric fraction \cite{Smith2009}, defined as: $(+)\text{-EF} = {N_+}/{(N_+ + N_-)}$ and $(-)\text{-EF} = {N_-}/{(N_+ + N_-)}$, where $N_+$ and $N_-$ refer to the number of $(+)$ or  $(-)$ particles within the region where the enantiomeric fraction is evaluated. The $(+)$-EF=70\% within $z \in [-0.330, 0.330]$ $\mu$m, and $(-)$-EF=92\% outside that region.

Lastly, we must comment on the practicality of this separation method. While it is true that it produces $F_{\rm chiral}>F_{\rm achiral}$ for a wide range of chiralities, this is not enough to ensure enantioseparation. One important aspect to overcome is the enantiomeric mixing produced by the Brownian motion of the particles. Because this method uses trapping chiral forces for separation, the Brownian motion must be overcome. This places a quite strong limit on the range of chiralities and radii of particles: in general, this method is adequate to separate big particles exhibiting small chirality. For instance, particles with r=100 nm $\kappa < 0.05$ would not be separable. 

In the case where $\Delta n = 0$, we can use another sorting method for absorbing chiral particles. As shown in the next subsection, this next method can effectively bypass the Brownian motion by use of longitudinal chiral forces without trapping potentials. As such, smaller particles with smaller chiralities can be effectively separated.

\subsection{Chiral pressure forces}

\begin{figure}[htb!]
    \centering
    \includegraphics[width=0.8\textwidth]{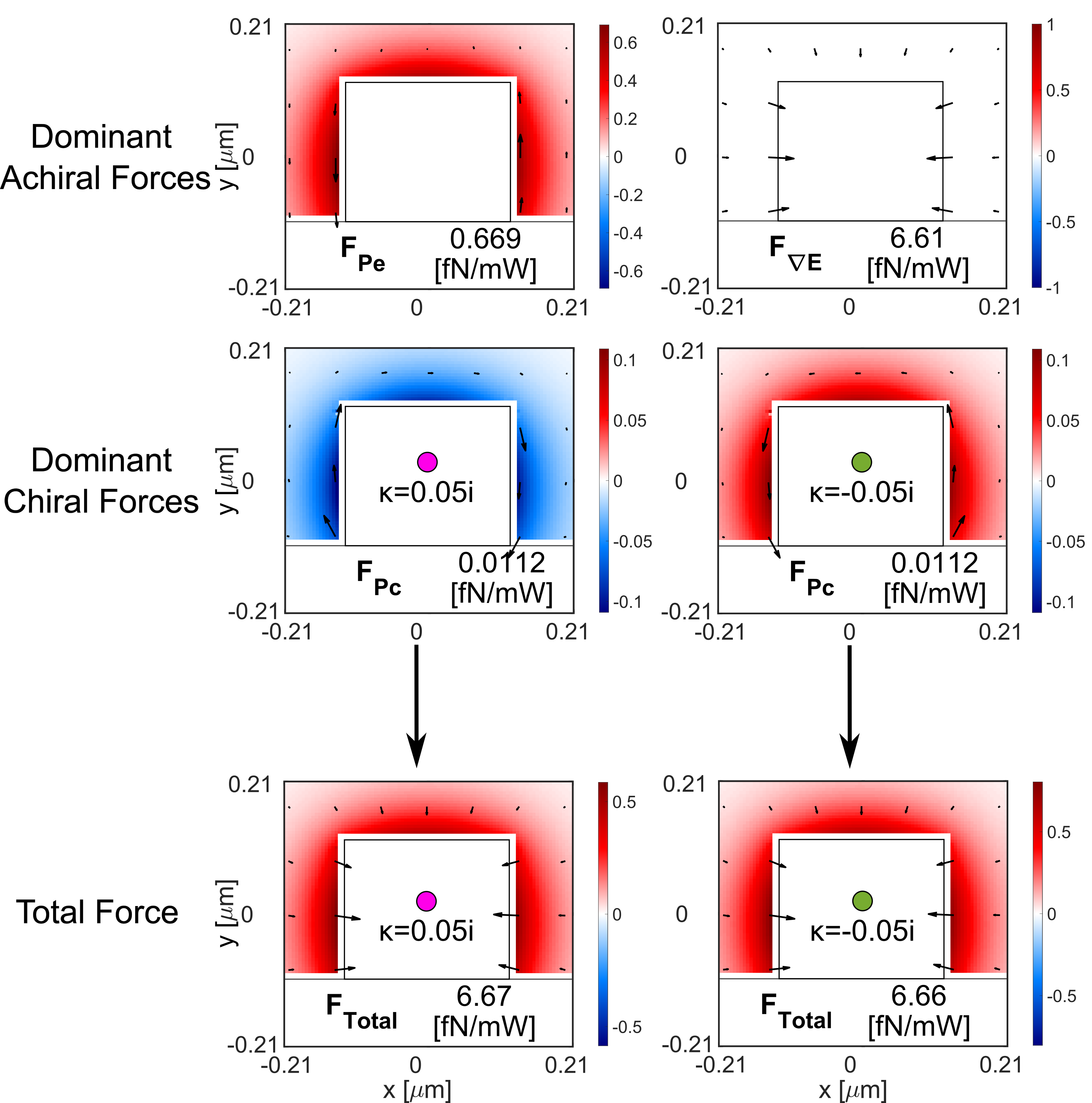}
    \caption{Optical forces acting on a chiral gold particle of $r=10$ nm and $\kappa = 0.05i$ throughout the cross-section of the degenerate waveguide system. We show the dominant achiral, and dominant chiral forces as well as the total force. The main difference between the total force experienced by both enantiomers is the maximum value along the longitudinal direction being 0.586 fN/mW for $\kappa = +0.05i$ and 0.804 fN/mW for $\kappa = -0.05i$; this difference enables the longitudinal separation. In the transversal plane, both enantiomers experience an achiral attractive force towards the waveguide core due to the electric density gradient.}
    \label{Forces10}
\end{figure}

\begin{figure}[t]
    \centering
    \includegraphics[width=0.95\textwidth]{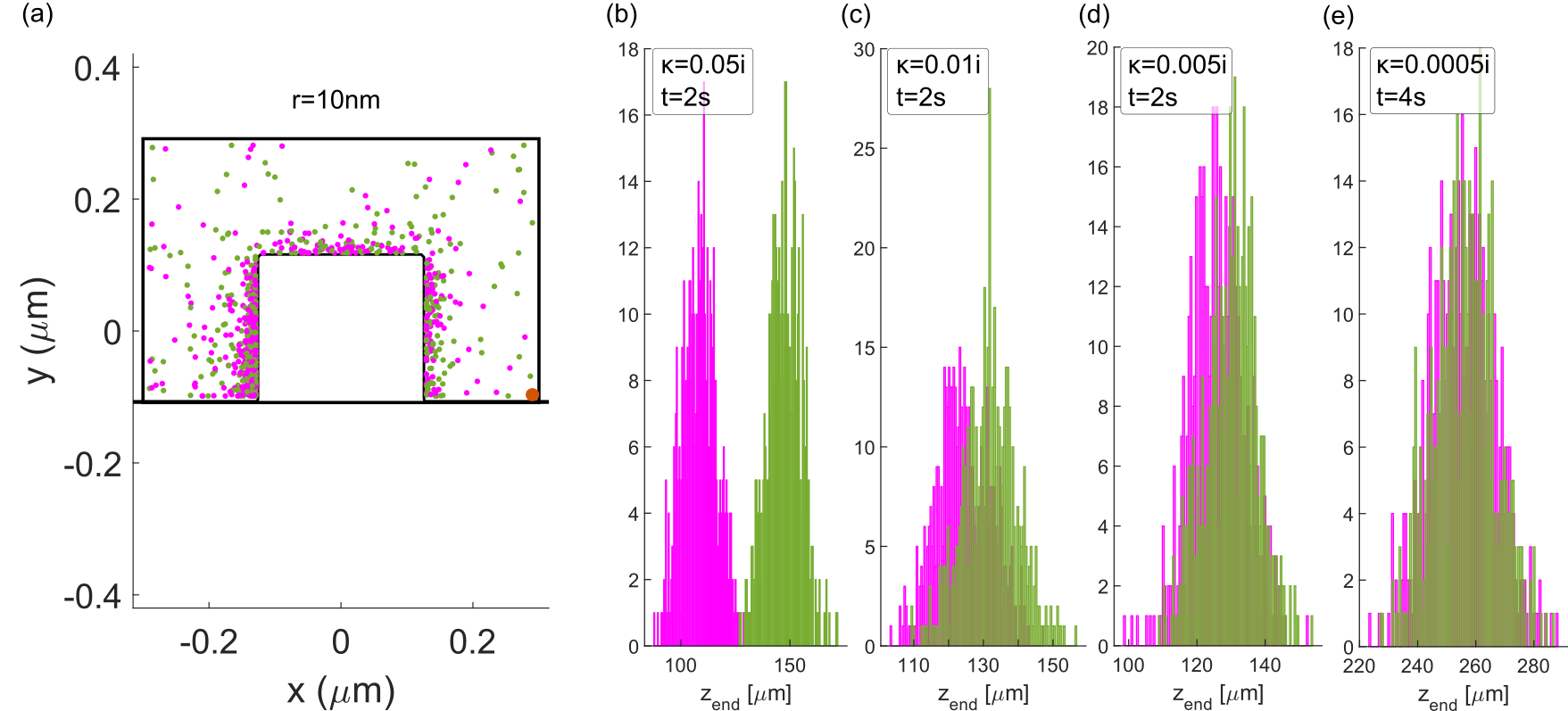}
    \caption{Result of the particle tracking simulations for chiral gold nanoparticles of 10 nm radius. (a) Cross-sectional view of the microchannel-waveguide system that represents the final position of both enantiomers (with $\kappa=\pm0.05i$, 500 particles each) within the $xy$-plane after the simulated time. The microchannel has 0.6 $\mu$m width and 0.4 $\mu$m thickness. Most particles accumulate close to the waveguide regardless of their chirality sign. The final $z$-position for both enantiomers of different $\kappa$ values are represented as histograms in (b) $\pm0.05i$, (c) $\pm0.01i$, (d) $\pm0.005i$ and (e) $\pm0.0005i$. A lesser degree of separation of the enantiomers distribution is obtained within 2 seconds for a lower chirality parameter.}
    \label{PartT10}
\end{figure}

We consider a strip waveguide made of a silicon nitride core (0.239 $\mu$m width $\times$ 0.217 $\mu$m thickness)
with refractive index $n\approx2.02$ on a SiO$_2$ substrate $(n=1.4468)$, surrounded by water ($n = 1.33$), operating at a wavelength of 633 nm. The optical mode is a combination of a quasi-TE and a quasi-TM mode that is delayed 90 degrees with respect to each other. For the chosen waveguide cross-section and wavelength, the quasi-TE and quasi-TM modes show degeneracy. Therefore, the combination results in a quasi-circularly polarized compound mode whose polarization is maintained along the waveguide length. The degeneracy condition ($n_{\rm TE}=n_{\rm TM}$) was found by sweeping the waveguide width for a fixed waveguide thickness and wavelength. The chiral particles to be studied in this system are assumed to be non-magnetic ($\mu_{\rm p} = 1$) gold spheres ($\varepsilon_{\rm p} = -11.753 + 1.2596i$). Four different chirality parameter values were studied $\kappa=0.05i$, $0.01i$, $0.005i$, and $0.0005i$, for two different radii: 10 nm and 50 nm. By considering a purely imaginary value for $\kappa$, we are implicitly assuming that the particle exhibits a maximum in its circular dichroism spectrum at the selected wavelength (633 nm).


The dominant achiral and chiral forces, as well as the total force for each enantiomer, are represented throughout the waveguide cross section in Fig. \ref{Forces10}, for a particle with $r=10$ nm and $\kappa=\pm0.05i$. The colormap represents the $z$-component of the forces and the arrowmap the transversal components. These force fields are maintained along the waveguide length because both the quasi-TE and quasi-TM modes are degenerate (in this case there is no beating pattern). In the transversal directions ($xy$-plane), the force field attracts both enantiomers towards the waveguide due to the dominant achiral electric energy density gradient force. In the longitudinal direction, the dominant force is the achiral electric pressure ($F^{ p_e}$). However, the opposite chiral pressure term ($F^{\real p_c}$)  with values $\sim\pm0.1$ fN/mW for opposite enantiomers, results in the $(+)$-enantiomer being pushed with a net force of $\sim0.5$ fN/mW, and the $(-)$-enantiomer being pushed with $\sim0.7$ fN/mW.

We have tested the sorting capability of this system by performing particle tracking simulations using the force field from Fig. \ref{Forces10} for a guided power of 100 mW.  For these simulations we have taken into account the Brownian motion as explained in \cite{Martinez-Romeu2024}. We obtained the final positions for 500 $(+)$-enantiomers and 500 $(-)$-enantiomers after a given amount of time. The initial (x,y) positions of the particles were randomized throughout the microchannel. 

Results for a particle of 10 nm radius in a microchannel 600nm wide and 400 nm thick are shown in Fig.\ref{PartT10}:  the final positions within the $xy$-plane are represented in (a), and the final positions along $z$ are shown in histogram plots for particle of different chirality parameter: (b) $\pm0.05i$, (c) $\pm0.01i$, (d) $\pm0.005i$, (e) $\pm0.0005i$. Particles of higher chirality achieve greater separation along the longitudinal direction than those of low chirality for an equal amount of time, as suggested by how separated the $z_{\rm end}-$position distributions for each enantiomer are. In order to get an estimation of the sorting time for particles with the chirality parameters in (b)-(e), we have made the following considerations. We consider the enantiomers are separated when the distance between the mean values of each distribution ($\overline{z}_{+}$ and $\overline{z}_{-}$) is at least larger than four times the average standard deviation of both clouds ($\sigma_{z}$), i.e. we calculated the time at which $|\overline{z}_{+}-\overline{z}_{-}|=4\sigma_{z}$. This calculation is explained in more detail in section \ref{app_time_evol} of the appendix. The time needed for obtaining separation for each case was estimated to be: (c) 23.5 s, (d) 93.8 s, (e) 9620 s = 2.7 hours.

\begin{figure}[t]
    \centering
    \includegraphics[width=0.8\textwidth]{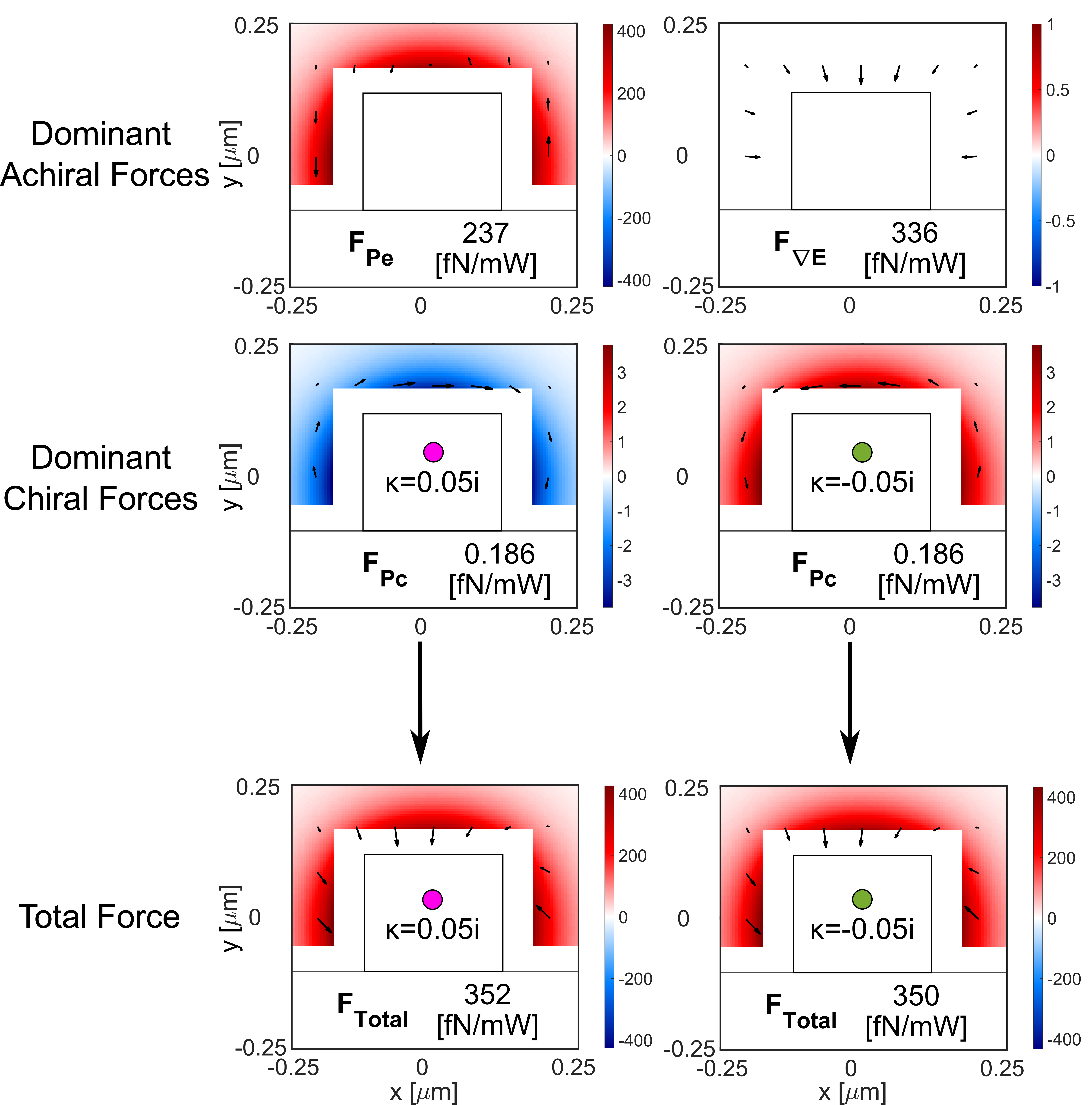}
    \caption{Optical forces acting on a chiral gold particle of $r=50$ nm and $\kappa = \pm0.05i$ throughout the cross-section of the degenerate waveguide system. We show the dominant achiral, the dominant chiral forces, and the total force. The main difference between the total force experienced by both enantiomers is the maximum value along the longitudinal direction being 426 fN/mW for $\kappa = +0.05i$ and 434 fN/mW for $\kappa = -0.05i$; this difference enables the longitudinal separation. In the transversal plane the combination of the Poynting force, the electric pressure, and the electric gradient will push the particles to the left corner of the waveguide.}
    \label{Forces50}
\end{figure}

\begin{figure}[t]
    \centering
    \includegraphics[width=0.95\textwidth]{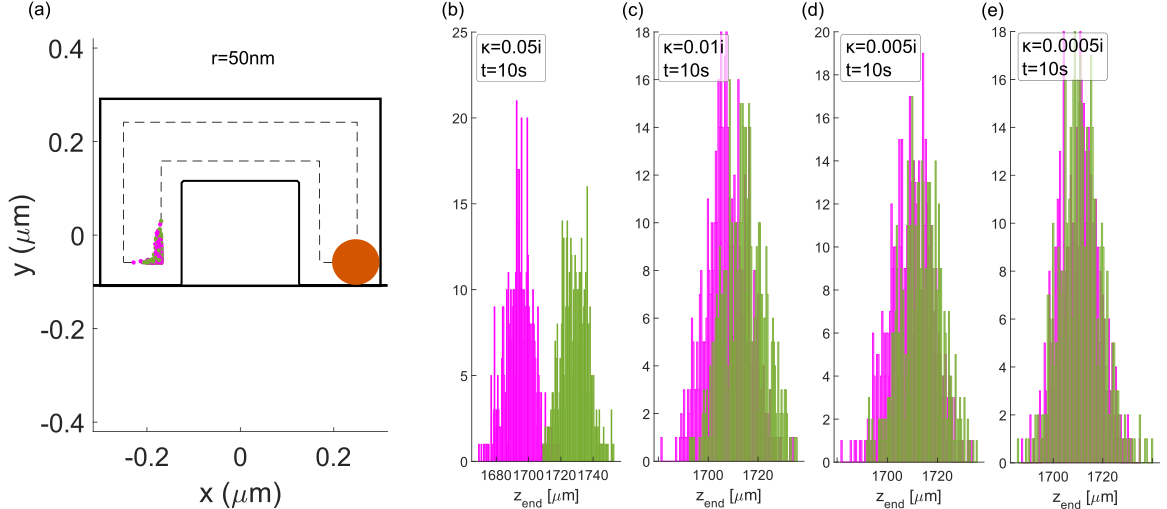}
    \caption{Result of the particle tracking simulations for chiral gold nanoparticles of 50 nm radius. (a) Cross-sectional view of the microchannel-waveguide system that represents the final position of both enantiomers (with $\kappa=\pm0.05i$, 500 particles each) within the $xy$-plane after 2 seconds. The microchannel has 0.6 $\mu$m width and 0.4 $\mu$m thickness. Most particles accumulate close to the waveguide regardless of their chirality sign. The final $z$-position after 10 seconds for both enantiomers of different $\kappa$ values are represented as histograms in (b) $\pm0.05i$, (c) $\pm0.01i$, (d) $\pm0.005i$ and (e) $\pm0.0005i$. A lesser degree of separation of the enantiomers distribution is obtained within 10 seconds for a lower chirality parameter.
    }
    \label{PartT50}
\end{figure}

We repeated the same study (forces and sorting capabilities of the system) for a 50 nm radius particle. The dominant optical forces over the cross-section of the system are shown in Fig. \ref{Forces50}. For this particle size, the dominant forces are the achiral electric energy density gradient and the electric radiation pressure forces. The resulting force field yields the particle tracking as shown in Fig. \ref{PartT50}, where the main difference with respect to the 10 nm radius particles is that there is an achiral orbital movement of the particles due to the $F^{ p_e}$ force which accumulates transversally both enantiomers on the left side of the waveguide. Yet, the different longitudinal force magnitude of $426$ fN/mW and $434$ fN/mW for $(+)$ and $(-)$ enantiomers enable the longitudinal sorting. The time needed to achieve $|\overline{z}_{+}-\overline{z}_{-}|=4\sigma_{z}$ separation was estimated to be: (c) 187 s, (d) 12.5 minutes, (e) 9620 s = 20.8 hours.


The enantiomeric fraction (EF) was obtained in a different manner than the sorting enabled by the sum of the TE mode and TM mode which produced chiral beating. It is explained in detail in section \ref{app_EF} of the appendix. The EF is calculated for each enantiomer in its correspondent zone separated by the medium point between the centers of the $z_{\rm end}$-distributions. The EF corresponding to a separation of $4\sigma_z$ between both distributions is: $(+)$-EF=97.72\% for $z < (\overline{z}_{+} + \overline{z}_{-})/2$, and $(-)$-EF=97.72\% for $z > (\overline{z}_{+} + \overline{z}_{-})/2$. Where $\overline{z}_{+}$ and $\overline{z}_{-}$ are the central points of the corresponding distributions. This value is the same for both 50 nm and 10 nm radius particles evaluated at the sorting time correspondent to the assumed separation condition: $|\overline{z}_{+}-\overline{z}_{-}|=4\sigma_{z}$. 

Finally, we show that our separation method can separate realistic chiralities. We compared the $g$-factor exhibited by our simulated chiral gold nanoparticles (obtained with chiral Mie theory \cite{Bohren1975} and shown in Table \ref{tab:g-factors}) to that of known chiral molecules reported in \cite{Mason63}. The 3$\beta$-Hydroxy-5$\alpha$-androstan-16-one(13) exhibits a $g$-factor of 0.175 similar to a chiral gold nanoparticle of $r=10$ nm and $\kappa=\pm0.05i$. On the other end, the aminoacid L-cystine shows a $g$-factor$=0.002$, similar to that of a chiral gold nanoparticle of $r=10$ nm and $\kappa=\pm0.0005i$. This shows that our chosen chirality parameters correspond to realistic molecular chiralities.


\begin{table}
    \centering
    \begin{tabular}{|c|c|c|c|}
    \hline
        Type of chiral particle & $\kappa$  & $g$-factor \\
        \hline
        \multirow{2}{*}{Gold nanosphere r=10 nm} & $\pm0.05i$ & 0.1600 \\
         & $\pm0.0005i$ & 0.0016 \\
         \hline
         \multirow{2}{*}{Gold nanosphere r=50 nm} & $\pm0.05i$ & 0.0480\\
         & $\pm0.0005i$ & 0.0005 \\
        \hline
        Molecule 3$\beta$-Hydroxy-5$\alpha$-androstan-16-one(13)  &  - & 0.1750 \cite{Mason63} \\
        \hline
        Molecule L-cystine &  -  &  0.0020 \cite{Mason63}\\
        
    \hline
    \end{tabular}
    \caption{Studied chiral gold nanosphere's $g$-factors and molecules' $g$-factors tabulated in \cite{Mason63}. }
    \label{tab:g-factors}
\end{table}

        

\section{Conclusions}

We have exhaustively studied the potential of exploiting chiral longitudinal forces in photonic integrated waveguides for sorting absorbing and non-absorbing chiral particles of realistically low chirality. The separation of enantiomers capabilities of such forces was confirmed with particle tracking simulations.

For non-absorbing particles, we have designed a waveguide where the guided mode polarization varies between right-handed elliptical polarization and left-handed elliptical polarization in a periodic pattern, whereas the electromagnetic energy density is maintained longitudinally. We have shown that in such systems the chiral longitudinal gradient forces dominate over the achiral longitudinal forces even for particles of low chirality. In particular, 100 nm-radius particles of $\kappa = \pm0.05$ can be separated within 1 second.

For absorbing particles, we have designed a waveguide where the guided quasi-TE and quasi-TM modes are degenerate to maintain a quasi-circular polarization over the length of the waveguide. This system enables the separation of enantiomers of arbitrarily low chirality as long as enough time is waited. We have shown that gold particles of $\kappa = \pm0.0005i$ and radius either $50$ nm or $10$ nm can be separated in 21 hours and 3 hours, respectively. Our results unveil the potential of photonic integrated waveguides to become a platform for enantiomeric sorting of a wide variety of nanoparticles and even molecules. 

Acknowledgments: The authors acknowledge funding from the European Commission under the CHIRALFORCE Pathfinder project (grant no. 101046961) and from Innovate UK Horizon Europe Guarantee (UKRI 10045438). A.M. acknowledges partial funding from the Conselleria de Educación, Universidades y Empleo under the NIRVANA Grant (PROMETEO Program, CIPROM/2022/14).

Data available in Zenodo repository at \cite{Data}.

\section{Appendix}

\subsection{Derivation of helicity density longitudinal variation in chirality change of combination of TE and TM modes}\label{app_derivation}
Let us consider a guided left circularly polarized mode (LCP) described by the following electric field $\vc{E}_{\rm LCP}=\vc{E}_{\rm TE}e^{in_{\rm TE}kz} + i\vc{E}_{\rm TM}e^{in_{\rm TM}kz}$ and magnetic field $\vc{H}_{\rm LCP}=\vc{H}_{\rm TE}e^{in_{\rm TE}kz} + i\vc{H}_{\rm TM}e^{in_{\rm TM}kz}$. The derivative of the helicity density along the propagation direction $z$ will be given by:
  
\begin{equation}
\begin{split}
     \frac{d\mathfrak{G}}{dz} = &  \frac{d}{dz}\left[ \frac{1}{2\omega c} \imaginary \left( \vc{E}_{\rm LCP} \cdot \vc{H}^*_{\rm LCP} \right)\right] = \frac{1}{2\omega c} \frac{d}{dz}     \Im\left(\vc{E}_{\rm LCP} \cdot \vc{H}^{*}_{\rm LCP}\right)= \\
    = & \frac{1}{4\omega c}\frac{d}{dz}\Im\left([\vc{E}_{\rm TE}e^{in_{\rm TE}kz} + i\vc{E}_{\rm TM}e^{in_{\rm TM}kz}]\cdot[\vc{H}^{*}_{\rm TE}e^{-in_{\rm TE}kz} - i\vc{H}^{*}_{\rm TM}e^{-in_{\rm TM}kz}]\right)= \\
       = & \frac{1}{4\omega c}\frac{d}{dz}\Im\left(\vc{E}_{\rm TE}\cdot\vc{H}^{*}_{\rm TE} + \vc{E}_{\rm TM}\cdot\vc{H}^{*}_{\rm TM}\right) +   \\
     + & \frac{1}{4\omega c}\frac{d}{dz}\Im\left(\vc{E}_{\rm TE}\cdot\vc{H}^{*}_{\rm TM}(-i)e^{ikz(n_{\rm TE}-n_{\rm TM})} + \vc{E}_{\rm TM}\cdot\vc{H}^{*}_{\rm TE}(i)e^{-ikz(n_{\rm TE}-n_{\rm TM} )}\right) \\  
 \end{split}
 \end{equation}

\noindent where the first summand of the last expression is zero as it does not depend on $z$. The second summand can be expressed in a compact form using $\Im(z) =\frac{1}{2i}(z-z^{*}) \;\; {\forall z\in \mathbb{C}} $ and defining $\psi=\vc{E}_{\rm TE}\cdot \vc{H}_{\rm TM}^*-\vc{E}_{\rm TM}^*\cdot \vc{H}_{\rm TE}$, and $\Delta n =n_{\rm TE} - n_{\rm TM}$:

\begin{equation}
\begin{split}
     \frac{d\mathfrak{G}}{dz} = & -\frac{1}{4\omega c} \frac{d}{dz} \left[\psi e^{ikz\Delta n} + \psi^*e^{-ikz\Delta n}\right] =\\  
     = & -\frac{1}{4\omega c} \frac{d}{dz} \left[\left|\psi\right| \frac{e^{i(kz\Delta n + \arg \psi)} + e^{-i(kz\Delta n + \arg \psi)}}{2} \right]= \\
     =& -\frac{1}{4\omega c} \frac{d}{dz} \left[ \left|\psi\right| \cos(kz\Delta n + \arg \psi) \right]\\
     =& \frac{1}{4\omega c}\left|\psi\right| k\Delta n \sin(k\Delta n z + \arg\psi)\\
\end{split}
\end{equation}

\subsection{Analysis of components of the chiral gradient force in parameter space}\label{app_force_analysis}

\begin{figure}[t]
    \centering
    \includegraphics[width=1\linewidth]{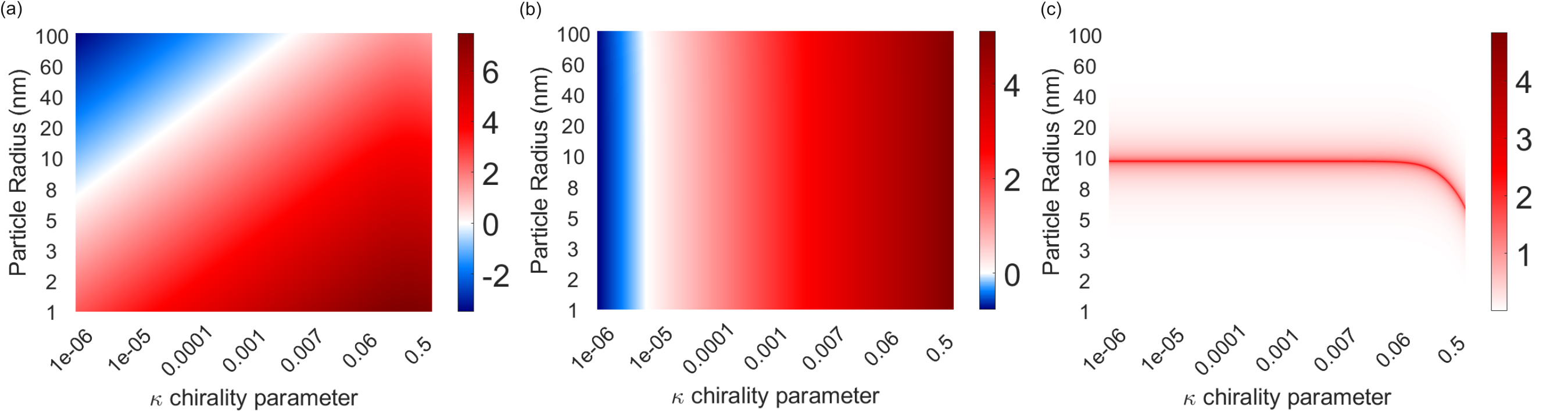}
    \caption{The colormaps represent different ratios of forces for the silicon waveguide, operating at 1310 nm wavelength. (a) $\log_{10}(|F^{\nabla \mathfrak{G}}_{z}/F^{\real\Pi}_{z}|)$. (b) $\log_{10}(|F^{\nabla \mathfrak{G}}_{z}/F^{\nabla E}_{z}|)$. (c) $\log_{10}(|(F^{\real\Pi}_{z}-F^{\nabla E}_{z}/F^{\real\Pi}_{z}+F^{\nabla E}_{z})|)$. }
    \label{fig:Comparative-Forces}
\end{figure}

Figure \ref{fig:Comparative-Forces}a shows that the relative strength (ratio) of the real Poynting vector force ($F^{\real\Pi}_z$) and the longitudinal helicity density gradient force ($F^{\nabla \mathfrak{G}}_{z}$) depends on the radius of the particle. This is because for non-absorbing particles the $F^{\real\Pi}_z$ grows with the product of two polarizabilities (particle volume squared), whereas $F^{\nabla \mathfrak{G}}_{z}$ grows linearly with the polarizability (particle volume). Thus, $F^{\nabla \mathfrak{G}}_{z}$ dominates over $F^{\real\Pi}_z$ in region II. Figure \ref{fig:Comparative-Forces}b shows that since both the achiral electric energy density gradient force ($F^{\nabla E}_{z}$) and $F^{\nabla \mathfrak{G}}_{z}$ depend linearly on the polarizability, their ratio does not depend on the particle radius. Thus, $F^{\nabla \mathfrak{G}}_{z}$ dominates over $F^{\nabla E}_{z}$ in region I. Figure \ref{fig:Comparative-Forces}c shows that both achiral forces, $F^{\real\Pi}_z$ and $F^{\nabla E}_{z}$, cancel each other for some specific radius (large ratio because the denominator is very small), which originates the distinctive line of minimum achiral force at $r=10$ nm in Fig. \ref{fig:Radii_kappa_sweep}a.

\subsection{Estimation of the separation time of enantiomers in a microchannel}\label{app_time_evol}
The movement of a particle in a fluid is governed by the combined action of the optical force (in our case), the drag force due to the viscosity of the fluid, and the stochastic Brownian motion. More details about the governing equation can be found in \cite{Martinez-Romeu2024}. The optical force makes the particle move in a deterministic manner with a terminal velocity $v_z$ that is proportional to the optical force and the mobility $\mathcal{M}$ of the particle in the fluid. The particle displacement along the $z-$axis due to the $z-$component of the optical force $F_z$ is given by $z_{\rm opt}$:
\begin{equation}
\begin{split}
    z_{\rm opt} = v_{z}t = \mathcal{M} F_z t 
    \label{eq:z_position}      
\end{split} 
\end{equation}
where the mobility is related to the viscosity of the fluid ($\eta$) and the radius of the spherical particle ($r$) according to $\mathcal{M}=1/(6\pi \eta r)$. In addition, the stochastic effect of the Brownian motion is to spread the possible positions around an average value.
Thus, the dispersion in position $z$ of a particle evolves in time as a normal distribution with a standard deviation given by:

\begin{equation}
    \sigma_{z}=\sqrt{2Dt}=\sqrt{2\mathcal{M}k_{B}Tt}
    \label{eq:z_sigma}
\end{equation}

Once given an initial distribution of $z-$position of particles that have started in the same initial $z$-position (assumed 0 here), we can model their time evolution by describing the movement of the average position $\bar{z}$ with Eq.~\ref{eq:z_position} and its dispersion $\sigma_z$ with Eq.~\ref{eq:z_sigma}. We refer to these position distributions as clouds. Each enantiomer, referred to as $(+)$ and $(-)$, experiences a different optical force due to the opposite sign of the chiral term, correspondingly $F^+ = F_{\rm achiral} + F_{\rm chiral}$ and $F^- = F_{\rm achiral} - F_{\rm chiral}$. Therefore, the terminal velocity $v_z$ is different for each enantiomer, but the spread due to brownian motion is the same $\sigma_z$ for both.

We have applied this modeling to the final distribution of positions resulting from the particle tracking simulations. This allows us to predict the time at which the cloud of positive and negative enantiomers will be separated, assuming a constant optical force acting on the particles.
We have expressed the desired separation between the two enantiomer clouds, with average positions $\bar{z}_+$ and $\bar{z}_-$, as a multiple of their dispersion, $s\sigma_{z}$, where $s$ can be any positive number. 
\begin{equation}
    |\bar{z}_{+}  -  \bar{z}_{-}| = s\sigma_{z}
\end{equation}
Note that both enantiomer clouds follow the same dispersion evolution, as both enantiomers have the same size. Therefore, by solving for $t$ and using Eq.\ref{eq:z_position} and Eq.\ref{eq:z_sigma}, we find the expression to estimate the separation time:
\begin{equation} \label{eq:time_estimation}
    t = \frac{s^2 2 k_B T}{\mathcal{M} \langle F^+_z-F^-_z \rangle ^2}=\frac{s^2 k_B T}{2\mathcal{M} \langle F_{\rm chiral,z}\rangle ^2}
\end{equation}
\noindent where we have used $\langle F^{+}_z-F^{-}_z\rangle=2\langle F_{\rm chiral,z}\rangle$.

\begin{figure}[t]
    \centering
    \includegraphics[width=0.8\textwidth]{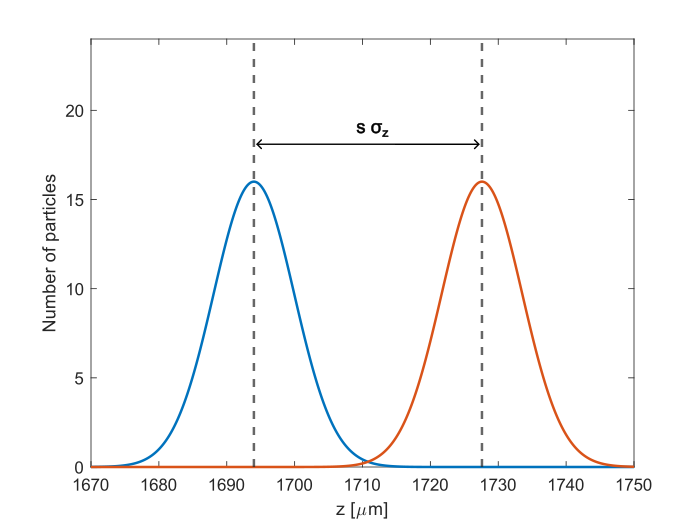}
    \caption{Example of separation of a racemic mixture where the difference in the mean position of the distributions is $4\sigma$. We have used the mean position of the radi 50 nm particles at 10 s and their $\sigma_{z}$.}
    \label{4sigma}
\end{figure}

Using these equations, it is possible to make inferences of separation times for any degree of separation (regulated by $s$). To make accurate predictions, we must take into account that the mobility of the particle will change due to its proximity to a wall from the microchannel or waveguide core. We must also consider that the magnitude of the force at each instant of time is in general smaller than the maximum force throughout the waveguide cross-section. We used the last positions from the tracking simulation to calculate the correction to $\mathcal{M}$ and the value for $\langle F_{\rm chiral,z}\rangle$. 

In our case, we cannot directly use Eq.~\ref{eq:z_position} for small chiralities to get the estimate of the average chiral force $\langle F_{\rm chiral,z}\rangle$ experienced by the particles from simulation. This is because the difference in positions is much smaller than the standard deviation for the simulated time. Instead, we first use the sum of the mean position after a simulation time $t_{\rm simulation}$ to get an estimate of the average achiral force $\langle F_{\rm achiral,z}\rangle$:
\begin{equation}
\begin{split}
     \frac{\bar{z}_{+}+\bar{z}_{-}}{2}=\frac{\mathcal{M} \langle F^{+}_z+F^{-}_z\rangle t}{2} = \;\;\;\;\;\;\;\;\;\;\;\;\;\\
    = \frac{\mathcal{M} \langle 2F_{\rm achiral,z} +F_{\rm chiral,z} -F_{\rm chiral,z} \rangle t}{2} =\mathcal{M} \langle F_{\rm achiral,z} \rangle t    
\end{split}
\end{equation}
\begin{equation} \label{eq:average_achiral}
    \langle F_{\rm achiral,z}\rangle=\frac{\bar{z}_{+}+\bar{z}_{-}}{2\mathcal{M}t_{\rm simulation}}
\end{equation}

Then, we assume that the strength ratio between the average chiral and achiral force $\langle F_{\rm chiral,z}\rangle / \langle F_{\rm achiral,z}\rangle$ is the same as the strength ratio between the maximum values throughout the cross section of the microchannel $|F^{max}_{\rm chiral,z}/F^{max}_{\rm achiral,z}|$. 
Thus, we estimate the average chiral force as: 
\begin{equation} \label{eq:chiral_achiral_ratio}
    \langle F_{\rm chiral,z}\rangle=\langle F_{\rm achiral,z}\rangle  \abs{\frac{F^{max}_{\rm chiral,z}}{F^{max}_{\rm achiral,z}}}
\end{equation}
With those, we can finally accurately estimate the times of separation for any degree of separation using Eq.~\ref{eq:average_achiral}, Eq.~\ref{eq:chiral_achiral_ratio} and Eq.~\ref{eq:time_estimation} in this order.

\subsection{Analytical calculation of enantiomeric fraction}\label{app_EF}
To calculate the enantiopurity of the separated racemic mixtures after the separating process we use the enantiomeric fraction, given by the expression \cite{Smith2009}:
\begin{equation}
    (+)\text{-EF}=\frac{N_{+}}{N_{+}+N_{-}}
\end{equation} 
We can do so analytically in the case of the absorbing particles. First, we must imagine the situation described by Fig. \ref{4sigma}, where both Gaussians have equal standard deviation $\sigma$ and their centers are separated by 4$\sigma$.

If we take the division line to be at the center of the two separated normal distributions, the position from the center of the left distribution (which we will define as the +) is $\mu_{+} + 2\sigma$ where $\mu_{+}$ is the center of the + cloud of enantiomers. This center, seen from the perspective of the other enantiomeric cloud (which we will call the - enantiomeric cloud) will be given as $\mu_{-} - 2\sigma$.

Now, because the enantiomeric clouds follow a Gaussian distribution we know that the number of positive enantiomer particles to the left of the center is given by: $N_{+} = N^{total}_{+} P_{+}(x \leq \mu_{+} + 2\sigma)$. Similarly, the number of - particles to the left of the center will be given by $N_{-} = N^{total}_{-} P_{-}(x \leq \mu_{-} - 2\sigma)$. We can use the expressions on the enantiomeric fraction and we get:
\begin{equation}
    (+)\text{-EF}=\frac{N^{total}_{+} P_{+}(x \leq \mu_{+} + 2\sigma)}{N^{total}_{+} P_{+}(x \leq \mu_{+} + 2\sigma)+N^{total}_{-} P_{-}(x \leq \mu_{-} - 2\sigma)}
\end{equation} 

In our case we consider $N^{total}_{+}=N^{total}_{-}$ which greatly simplifies the enantiomeric fraction expression to:
\begin{equation}
    (+)\text{-EF}=\frac{ P_{+}(x \leq \mu_{+} + 2\sigma)}{P_{+}(x \leq \mu_{+} + 2\sigma)+P_{-}(x \leq \mu_{-} - 2\sigma)}
\end{equation} 

We can use that $P_{-}(x \leq \mu_{-} - 2\sigma)=1-P_{-}(x \geq \mu_{-} - 2\sigma)$ and furthermore $P_{-}(x \geq \mu_{-} - 2\sigma)=P_{+}(x \leq \mu_{+} + 2\sigma)$ due to the probabilities of both enantiomers following the same Gaussian distribution. So finally, we get that: 

\begin{equation}
    (+)\text{-EF}=\frac{ P_{+}(x \leq \mu_{+} + 2\sigma)}{P_{+}(x \leq \mu_{+} + 2\sigma)+1-P_{+}(x \leq \mu_{+} + 2\sigma)}= P_{+}(x \leq \mu_{+} + 2\sigma)
\end{equation} 

By symmetry arguments, the $(-)$ enantiomer will yield the same EF on its right side. 
\begin{equation}
    (-)\text{-EF}= P_{-}(x \geq \mu_{-} - 2\sigma)
\end{equation} 

An important consideration to take from this derivation is that we can change the enantiomeric fraction simply by choosing the $s$ parameter (defined by being the separation between the centers of the distributions). 
\begin{equation}
    (+)\text{-EF}= P_{+}(x \leq \mu_{+} + \frac{s}{2}\sigma)
\end{equation} 
Different $s$ parameters will mean different waiting times due to Eq. \ref{eq:time_estimation} so we can adjust the waiting time to obtain any enantiomeric purity.

\bibliography{sample}
     
\end{document}